\documentclass[oribibl]{llncs}
\usepackage{url}
\usepackage{amsmath, amssymb}
\usepackage{graphicx}
\usepackage{capt-of}

\title{Document Retrieval for Large Scale Content Analysis using Contextualized Dictionaries}

\author{Gregor Wiedemann
\and Andreas Niekler}

\institute{University of Leipzig \\
   Augustusplatz 10 \\
   04109 Leipzig, Germany \\
   \url{gregor.wiedemann@uni-leipzig.de}, \url{aniekler@informatik.uni-leipzig.de} }

\begin{document}
\maketitle
\begin{abstract}
  This paper presents a procedure to retrieve subsets of relevant documents from large text collections for Content Analysis, e.g. in social sciences. Document retrieval for this purpose needs to take account of the fact that analysts often cannot describe their research objective with a small set of key terms, especially when dealing with theoretical or rather abstract research interests. Instead, it is much easier to define a set of paradigmatic documents which reflect topics of interest as well as targeted manner of speech. Thus, in contrast to classic information retrieval tasks we employ manually compiled collections of reference documents to compose large queries of several hundred key terms, called dictionaries. We extract dictionaries via Topic Models and also use co-occurrence data from reference collections. Evaluations show that the procedure improves retrieval results for this purpose compared to alternative methods of key term extraction as well as neglecting co-occurrence data.
\end{abstract}

\section{Introduction}

Due to the vastly growing availability of (retro-)digitized large scale text
corpora computer-assisted Content Analysis (CA) is of increasing interest for various
disciplines and applications ranging from social sciences to business
intelligence. When exploring large corpora analysts are confronted with the
problem of selecting relevant documents for qualitative investigation and
further quantitative analysis. Standard tasks in Information Retrieval (IR) usually rely on small sets of concrete key terms for querying a collection. Highly abstract research interests in CA often cannot describe research objectives by such small queries. For example, analysis on the war in Iraq certainly can query for \textit{Iraq AND war}. But, what would be a reasonable query for \textit{documents containing neoliberal justifications of politics}? 

To meet special requirements of CA retrieval tasks we propose a procedure where a query is not (primarily) based on single terms, but on a set of reference documents. Compared to the problem of determining concrete key terms for a query it is rather easy for analysts to manually compile a collection of `paradigmatic' documents which reflect topics or manner of speech matching their research objective. Retrieval with such a reference collection is then performed in three steps:

\begin{enumerate}
\item Extraction of a set of key terms from the reference collection, called dictionary. Terms in the dictionary are ranked by weight to reflect difference in their importance for describing an analysis objective.
\item Extraction of co-occurrence data from the reference collection as well as from an additional generic corpus representative of a given language.
\item Scoring relevancy of each document on the basis of dictionary and co-occurrence data to create a ranked list of documents.
\end{enumerate}

\noindent
\textbf{Related work:} Using documents as queries is a consequent idea within the Vector Space Model (VSM) of IR where key term queries are modeled as document vectors for comparison with documents in the target collection~\cite{Salton1975}. Our approach extends the standard VSM approach by additionally employing of shared meanings of topic defining terms captured by co-occurrence data. Co-occurrence data has been used in standard IR tasks for term weighting as well as query expansion with mixed results (e.g. see~\cite{Peat.1991, Billhardt.2000, VanRijsbergen.1977, Wong1985}). These applications differ from our approach as they want to deal with unequal importance of terms in a single query due to term correlations in natural language. The method presented in this paper does not weight semantically dependent query terms by co-occurrence information globally. Instead, in CA analysts are often interested in certain aspects of meaning of specific terms. Following the distributional semantics hypothesis~\cite{firth1957} meaning may be captured by contexts better than just by isolated terms. Therefore, we score relevancy based on similarity of individual contexts of single query terms in sentences of the target documents compared to observed contexts from the reference collection.

The paper is organized in 5 sections. After having clarified our motivation the next section presents our approach of semi-supervised dictionary extraction with the help of statistical Topic Models. The third part explains how to utilize ranked dictionaries together with co-occurrence data for document retrieval. Section 4 briefly introduces an example application of this procedure on a target collection of German newspaper articles, followed by an evaluation of the approach.

\section{Dictionary Extraction with Topic Models}

The generation and usage of dictionaries is an important part of quantitative
CA procedures \cite{krippendorff_content_2004}. Dictionaries provide the basis of code books and category systems within CA studies. As we want to exploit dictionaries for document retrieval we suggest a procedure of semi-supervised term extraction for dictionary creation. For this we apply a statistical Topic Model on a collection of paradigmatic documents. Reference documents should be selected carefully by the analysts in consideration of representing domain knowledge or specific language use of interest. The resulting list may be compared to domain term extraction based on reference corpus comparison \cite{drouin_detection_2004} or tf/idf weighting of words. These calculate `keyness' of terms isolated from each other. In contrast to those the Topic Model based approach takes account of the fact that terms do not occur independently of each other. 

\textbf{Statistical Topic Models: } Topic Models are a set of statistical models for unsupervised extraction of latent semantic structures from document collections. (\cite{blei_latent_2003} first introduced \textit{Latent Dirichlet Allocation}). They generate results where underlying latent variables with $K$ dimensions
are extracted from a document collection. Those variables represent
distributions over words $\phi_{\cdot, k} = p(\mathbf{w} \vert z_k),
(k=1,\ldots,K)$ representing their alignment to a semantically coherent group which may be interpreted as topic. Words
$\mathbf{w}$ with a high probability $p(\mathbf{w} \vert z_k)$ in a topic $k$ represent its determining terms and allow for interpretation of the meaning of an underlying thematic composition.

\textbf{Extracting Dictionaries from Topic Models:}
\label{subsec:dict}	
Distributions $p(\mathbf{w} \vert z_k)$ represent a simplification of the collection content as a composition of topics. We can assume that highly probable words play an important role in the semantics of the whole corpus. Therefore, those words may be used to compile a dictionary of key terms within that collection. We define the weight of a term in the dictionary by the sum of its probability values in each topic.

In comparison to term frequency counts in a collection the probability weight of a
term in a corpus represents a word's contribution to a certain context. Even if a
context has relatively low evidence in the data because of a low frequency a 
term can have high probability $p(w_n \vert z_k)$ within a topic.
We don't want to overly bias the ranks in the dictionary with very improbable topics and their words---e.g. the high probability of the top terms in topics of low probability would be ranked almost equal to the top words in highly probable topics. Therefore, we need to normalize the term probabilities $p(\mathbf{w} \vert z_k)$
according to either their topic's probability $p(z_k)$ or their
term frequency $\text{tf}(w_n)$ within the corpus. We decide to normalize each terms probability in a certain topic with its term frequency, but to use log frequency to dampen the effect of high frequency terms. The final weight of a term in the dictionary is determined by 
	
	\begin{equation}
  		\text{tw}_n = \log(\text{tf}(w_n)) \sum_{k=1}^K p(w_n \vert z_k) 
	\end{equation}
	
where $K$ is the number of topics, $\text{tf}(\cdot)$ the term frequency, and $w_n$
the dictionary term. Descended sorting of term weights $\mathbf{tw}$ results in a list of ranked words which can be cut to a certain length $N$.

\textbf{Analyst supervision}: Within a Topic Model usually topics with undesired content can be identified. A few
topics normally capture rather syntactic information of the collection representing
co-occurring functional words in a corpus \cite{Alsumait_2009}. Other
topics, although capturing semantic structure, may be considered as irrelevant by the analyst with a view to her/his research interest. In contrast
to other methods of key term extraction the Topic Model approach allows to exclude those
unwanted semantic clusters. Before calculation of the weights of terms one has to identify these topics which do not represent meaningful structures and to exclude them from the set of $K$ topics. This is an important step for the analyst to exercise influence on the so far unsupervised dictionary creation process and a clear advantage over other methods of key term extraction.

\section{Retrieval with dictionaries}

Dictionaries can be employed as filters in IR systems reducing general collections to sub collections containing sets of documents of interest for further analysis. Using a dictionary of ranked terms for IR might be translated in the standard Vector Space Model (VSM) approach in combination with `term boosting'. In this approach prior knowledge of unequal importance of terms is incorporated into  query processing via factors for each term. A VSM-based scoring function can be computed for a document $d$ and a dictionary as query $q$ as follows:


	\begin{equation}
  		\text{score}_\text{VSM}(q,d) = \sum_{w \in q} \text{tf}(w,d) \cdot \text{boost}(w) \cdot \text{norm}(d)
	\end{equation}

Usually IR weightings consider \textit{inverse document frequency} of a term as a relevant factor. As the dictionary ranking is derived from Topic Models, information comparable to document frequency has indirectly already been taken into account. We skip the $idf$ factor for each term in favor of our own dictionary weight. Rank information from the dictionary can be translated into a boosting factor for the scoring function. We suggest a factor ranging between 0 and 1

	\begin{equation}
  		\text{boost}(w) = \frac{1}{\sqrt{\text{rank}(w)}}
	\end{equation}

for each term $w$ which reflects that the most prominent terms in a dictionary of $N$
terms are of high relevancy for the retrieval process while terms located nearer
to the end of the list are of more equal importance. 

To address the problem of document length normalization and identify relevant documents of all possible length we utilize pivoted unique normalization as introduced in \cite{Singhal}. Pivotal length normalization slightly lowers relevancy scores for shorter documents of a collection $D$ and consequently lifts the score for documents after a pivotal value determined by the average
document length. The normalization factor for each document is computed by

	\begin{equation}
  		\text{norm}(d) = \frac{1}{\sqrt{(1 - \text{slope}) \cdot \text{pivot} + \text{slope} \cdot | U_d | }}
	\end{equation}
	
where $U_d$ represents the set of unique terms occurring in document $d$ and $pivot$ is computed by $\text{pivot} = \frac{1}{|D|} \cdot \sum_{d \in D} |U_d|$.

When evaluation data is available, the value for \textit{slope} might be
optimized for each collection. Lacking a gold standard for our
retrieval task we set \textit{slope} to 0.7 which has proven to be a reasonable choice for
retrieval optimization in various document collections \cite{Singhal}. Further, the \textit{tf} factor should reflect on the importance of an individual term relative to the average frequency of unique terms within a document: $avgtf(d) = \frac{1}{|U_d|} \cdot \sum_{t \in U_d} tf(t,d)$. Thus, the final scoring formula yields a document ranking for the entire collection:

	\begin{equation}
	\label{eq:unigram}
	\text{score}_\text{dict}(q,d) = \sum_{w \in q} \frac{1 + \log(\text{tf(w,d)})}{1 + \log(\text{avgtf}(d))} \cdot \text{boost}(w) \cdot \text{norm}(d)
	\end{equation}

\textbf{Contextualizing dictionaries: }
\label{subsec:coocs}
The approach described above yields useful results when looking for documents which can be described by a larger set of key terms.
When it comes to more abstract research interests, however, which aim to identify certain meanings of terms or specific language use, isolated observation of terms may not be sufficient. Fortunately the approach described above can be augmented with co-occurrence data from the reference collection to judge on relevancy of occurrence of a single key term in our target document. This helps not only to disambiguate different actual meanings of a term, but also reflects the specific usage of terms in the reference collection.
Therefore we compute patterns of significant co-occurrences of the $N$ terms
in our dictionary with each other resulting in an $N \times N$ matrix $C$. Co-occurrences are observed in a
\textit{sentence} window. Statistical significance of a co-occurrence is calculated by the Dice measure which is the fraction of the count of all sentences containing term $a$ and term $b$ over the sum of all sentences containing each single term:

	\begin{equation}
	\text{dice}(a,b) = \frac{2 n_{ab}}{n_a + n_b}
	\end{equation}

We decided for this measure instead of more sophisticated co-occurrence significance tests like Log Likelihood \cite{dunning_accurate_1993} because it also reflects syntagmatic relations of terms in language relatively well \cite{Bordag_2008}, but, more important its values range is between 0 and 1 which makes measurements over different corpora comparable. This is useful for dealing with an unwanted effect we experienced when experimenting with co-occurrence data to improve our retrieval mechanism. Co-occurrences of terms in the sentences of a reference collection may reflect characteristics in language use of the included documents. However, certain co-occurrence patterns may reflect general regularities of language not specific to a collection of a certain domain (e.g. strong correlations between the occurrence of \textit{parents} and \textit{children} or multi word units like \textit{United States} in one sentence). Applying co-occurrence data to IR scoring tends to favor documents where many of those common language regularities can be observed.

Instead of using the co-occurrence matrix $C$ solely based on the reference collection we `filter' the co-occurrence data by subtracting a co-occurrence matrix based on a second, randomly composed reference corpus.\footnote{Suitable corpora for this purpose, such as the ones provided by the ``Leipzig Corpora Collection'' \cite{BHQ07} which is carefully maintained by computational linguists, may be seen as representative of common language characteristics not specific to a certain domain or topic.} A second $N \times N$ matrix $D$ of co-occurrences is computed from such a generic corpus. The subtraction of D from C delivers a matrix $C'$ reflecting the divergence of co-occurrence patterns in the reference collection compared to common language: $C' = \text{max}(C - D, 0)$.\footnote{\textit{max} asserts that all negative values in $C'$ (representing terms occurring less together in sentences of the reference collection than in sentences of common language) are set to zero.}
Values for common combinations of terms (\textit{United States}) in $C'$ are significantly lowered while combinations specific to the reference collection remain largely constant.

\textbf{Using sentence co-occurrences: }
To exploit co-occurrence data for IR the scoring function in eq. \ref{eq:unigram} has to be reformulated to incorporate a similarity measure between a co-occurrence vector profile of each term $w$ in the dictionary and each sentence $s$ in the to-be-scored-document $d$. Instead of using just term frequency we add information on contextual similarity:

	\begin{equation}
		\label{fn:tfsim}
  		tfsim(w,d) = \sum_{s \in d} \sum_{w \in s} \text{tf}(w,s) + \alpha \cdot \frac{\vec{s} \cdot C'_{\cdot, w}}{\|\vec{s}\| \cdot \|C'_{\cdot, w}\|}
	\end{equation}

The frequency of $w$ within a sentence (which usually equals 1) is incremented by the cosine similarity between sentence vector $\vec{s}$ (sparse vector of length $N$ indicating occurrence of dictionary terms in $s$) and the dictionary context vector for $w$ out of $C'$.
This measure rewards the relevancy score, if the target sentence and the 
reference term $w$ share common contexts. In case they share no common context
$tfsim$ is equal to $tf$. 

Because term frequency and cosine similarity differ widely in their range the influence of the similarity on the scoring needs to be controlled by a parameter $\alpha$. If $\alpha$ is set to zero $tfsim$ replicates simple term frequency counts. Values for $\alpha$ higher than $0$ yield a mixing of unigram matching and contextual information for the relevancy score. Optimal values for $\alpha$ can be calculated by our evaluation method (see section \ref{sec:eval}). The context-sensitive score is computed as follows:

	\begin{equation}
	\label{fn:coocmix}
	\text{score}_\text{context}(q,d) = \sum_{w \in q} \frac{1 + \log(\text{tfsim(w,d)})}{1 + \log(\text{avgtf}(d))} \cdot \text{boost}(w) \cdot \text{norm}(d)
	\end{equation}

\section{Example}

The procedure presented above is applied to a political science research task performed as part of a German research project. The project aims at analyzing the influence of neoliberal ideas on domestic politics by studying the discourse in public media. In our example application we identify documents which (potentially) contain neoliberal argumentation in a collection of 101.032 newspaper articles of the German magazine DIE ZEIT (volumes 2000--2009). This example is also used for evaluation of the method in section \ref{sec:eval}.

To retrieve documents of interest for further analysis political scientists compiled a reference corpus consisting of 36 German books and journal articles written by self-confessed neoliberal theorists (e.g. Milton Friedman). In this reference corpus sentence boundaries are detected and tokens are lemmatized. A topic model based on the Pitman-Yor Process \cite{teh_hierarchical_2006} is calculated. For this process all paragraphs of books and articles in the collection were treated as single `documents' for modeling. In the resulting 23 topics of this model we can identify a very large topic containing
only English words (originating mostly from bibliographies in the reference collection) 
which have been clustered by the topic model process. Since this 
topic does not represent meaningful semantics for the analysis analysts could exclude it for the process of dictionary extraction. Using the process described in Section \ref{subsec:dict} a dictionary of 500 key terms is extracted from the reference collection. Co-occurrence patterns of these 500 terms are extracted according to Section \ref{subsec:coocs}.

This contextualized dictionary is then used to query the target corpus of DIE ZEIT newspaper articles. The 2,000 highest ranked articles are retrieved and used as starting point for further qualitative and quantitative analysis procedures by the analysts. 

\section{Evaluation}
\label{sec:eval}

Determining a large set of key terms from a reference collection and extracting its co-occurrence profiles to compose a ``query'' is an essential step in the proposed retrieval mechanism to meet requirements of content analysts. Due to this, standard approaches of IR evaluation \cite{sanderson_best_2009} are not applicable. There are no test collections like the TREC datasets~\cite{voorhees_trec_2005} regarding such type of retrieval task. In order to evaluate our method we therefore follow two approaches:

\begin{enumerate}
  \item Generating a quasi-gold standard of \textit{pseudorelevant documents} to show performance improvements through the use of co-occurrence data as well as key term extraction via topic models,
  \item Judging on the overall validity with \textit{precision at k} evaluation in our example retrieval performed by domain experts.
\end{enumerate}

\noindent
\textbf{Generating pseudorels:} Due to the lack of proper gold standards for our special retrieval task we create a custom evaluation set to evaluate our approach. Therefore, we use a strategy of data fusion. This strategy merges results, namely ranked lists of documents, of multiple retrieval systems to a set of pseudorelevant documents. These \textit{pseudorels} can be used as an automatically generated quasi-gold standard. The approach proposed by \cite{nuray2006automatic} shows that merging results of different retrieval systems by the Condorcet method generates pseudorels which produce evaluation results of IR systems that highly correlate with the TREC testset rankings of IR systems. Although it is hard to judge on the overall absolute performance of a system, the procedure allows for relative judgment between tested systems. We employ this strategy in a two-fold manner:
\begin{enumerate}
  \item Optimizing parameter $\alpha$ for the best mix of unigram / co-occurrence matches
  \item Deciding whether using topic models improves retrieval results over a simple tf/idf measure for key term extraction.
\end{enumerate}

\noindent
For the latter a second dictionary is created on the basis of the highest 500 tf/idf measures for each term in the reference collection. In (\ref{fn:tfsim}) we set the influence of the context similarity within a sentence by the parameter $\alpha$. For the evaluation we will
treat every setting for this parameter as a different retrieval system in order
to artificially create a large set of different systems. We vary the parameter
$\alpha$ by steps of 2 in a range of $[0,30]$. Furthermore, a system is added where just context similarities contribute to the ranking by setting $\alpha=1$ and leaving out $tf(w,s)$ in (\ref{fn:tfsim}). Each of the 17 `systems' produces a ranked list of 2000 highest-scored documents using the contextualized dictionary created by the topic model approach.
Another set of retrieval systems is created in the same manner, but using the dictionary which was extracted with the tf/idf measure. Both sets together provide 34 lists of ranked retrieval results allowing for a comparison between the represented systems. For this, results of the systems need to be merged to a set of pseudorelevant documents by the following procedure taken from \cite{nuray2006automatic}:

\begin{enumerate}
  \item Selecting a set of the most biased systems:
	As we want to compare results in two dimensions (i. use of co-occurrence data, ii. key term extraction procedure) we select 4 from our 34 `systems': The system which neglects co-occurrence data ($\alpha=0$) and the system which solely relies on co-occurrence data---both combined with a dictionary generated by Topic Model or tf/idf.
  \item Select the top documents of each of the most biased systems as candidates for pseudorels. For each retrieved document in each system a norm weight can be computed and summed up over all systems: $n_d = \sum\nolimits_{s=1}^S m_s/i_{d,s}$, with system $s$, $m_i$ number of ranks in system $s$ and $i_{d,s}$ rank of document $d$ in system $s$. Figure ~\ref{fig:mean_rank} shows the sorted values of norm weights $n_d$ for the best 100 documents in our example. The first documents rank very high in most of the tested systems. Documents with lower $n_d$ ranks are retrieved only by a few of the tested systems and, thus, are considered not to be good candidates for pseudorels. We select the top 50 documents of each system.
  \item Rank the candidates for pseudorels by using the Condorcet method which relies on counting wins and losses of direct comparison of document rankings within the most biased systems.
  \item The selection of pseudorels should reflect only those documents which yield a robust ranking within the Condorcet method. Figure ~\ref{fig:wins_rank} shows
  that roughly half of the documents of our example dataset have distinguishable values. Others produce more equal amounts of wins. Following \cite{nuray2006automatic} we define the top 50\% of the Condorcet ranked list as pseudorels. 
\end{enumerate}

\begin{table}[t]
\begin{minipage}[t]{.475\textwidth }%
\centering

\includegraphics[width=1\textwidth]{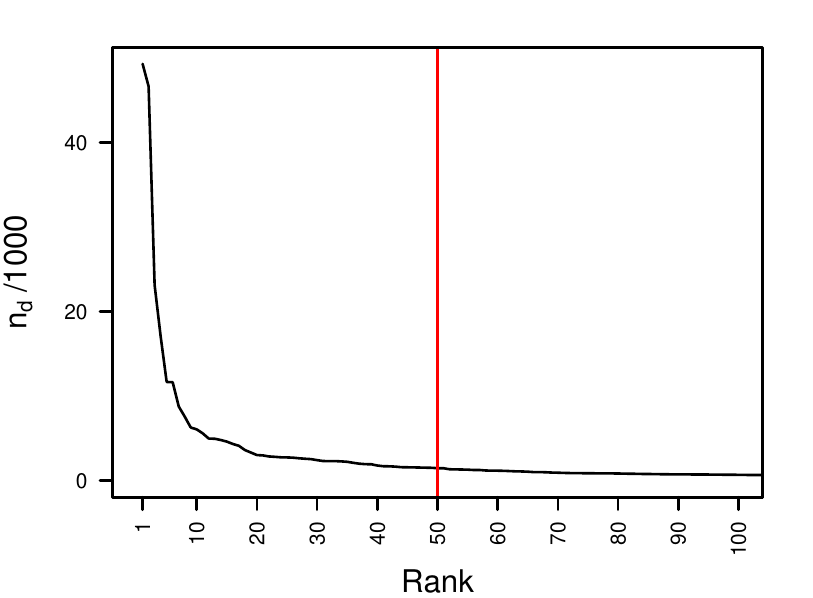}
\captionof{figure}{Plot of the accumulated values $n_d$ of all documents from
each retrieval system (first 100 ranks)}
\label{fig:mean_rank}
\end{minipage}%
\hspace{.05\textwidth}
\begin{minipage}[t]{.475\textwidth}
\centering
\includegraphics[width=1\textwidth]{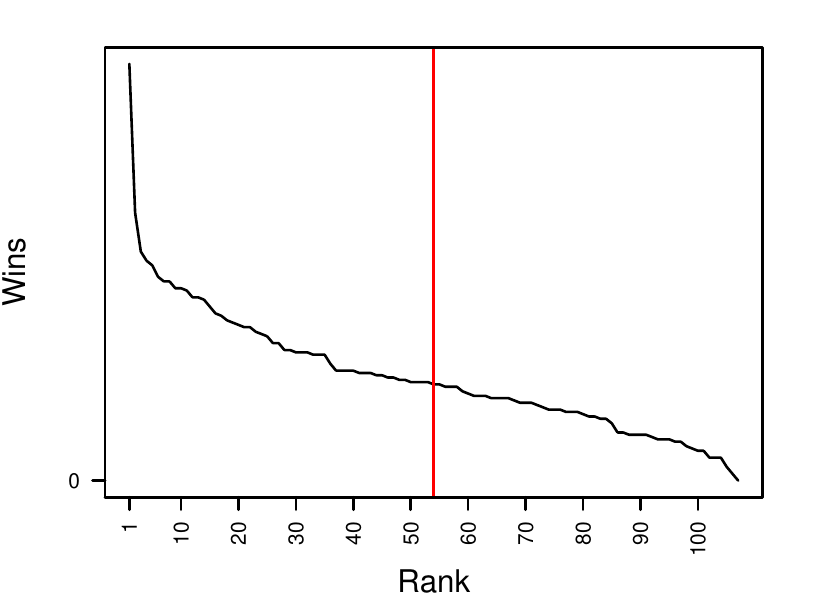}
\captionof{figure}{Plot of the wins of the documents found by the Condorcet
method (values have been sorted).}
\label{fig:wins_rank}
\end{minipage}

\end{table}

\noindent
\textbf{Mean average precision:} With this procedure a list of 54 documents is selected as pseudorels for evaluating the example application. Performance of each of the 34 retrieval systems is measured by utilizing the mean average precision (MAP). Since we only have one result e.g. one query for each system MAP is used as
$MAP = {1 \over R} \sum_{k=1}^{R} P(R[k]),$
with the number of relevant documents $R$ and $P(R[k])$ as precision
within the ranking of a system up to the document $R[k]$. 

The best performance of all tested systems is achieved by the system which mixes unigram and co-occurrence matching with parameter $\alpha=14$. Table \ref{ta:resMAP} displays MAP values for the two retrieval results based solely on unigram or context matching and the best performing mixed approach. Furthermore, it contrasts systems based on the tf/idf dictionary with systems based on the Topic Model dictionary. The results indicate that systems based on a Topic Model approach perform indeed better than systems using tf/idf for term extraction if co-occurrence data is used. This is not surprising considering the fact that Topic Models are based on co-occurrences as well. Our dictionary extraction method preserves this information in contrast to the independence assumption underlying tf/idf. Overall, the evaluation shows that using context information outperforms the base line approach without context information for our retrieval purpose. Nonetheless, because the evaluation method uses pseudorels care has to be taken in interpreting absolute MAP values of the systems.

\begin{table}[t]
\begin{minipage}[t]{.475\textwidth }%
\centering

\begin{center}
\begin{tabular}{ l | l | l}
\hline
  \textbf{Retrieval System} & \textbf{tf/idf} & \textbf{TM} \\
  \hline
  Unigram & 0.732 & 0.714 \\
  Co-occcurence & 0.657 & 0.723\\
  $\alpha$-mix & 0.823 & \textbf{0.861}\\ 
      & $\alpha=6$ & $\alpha=14$\\
\hline
\end{tabular}
\end{center}
\captionof{table}{MAP evaluation of compared retrieval systems.}
\label{ta:resMAP}

\end{minipage}%
\hspace{.05\textwidth}
\begin{minipage}[t]{.475\textwidth}
\centering

 \begin{center} 
 
 \begin{tabular}{l|l|l}
 \hline
 \textbf{Rank}  & $\mathbf{\alpha=0}$ & $\mathbf{\alpha=14}$ \\
 \hline
 1-10 & 0.9      & 1.0 \\
 101-110 & 0.7   & 1.0 \\
 501-510 & 0.8   & 0.9 \\
 1001-1010 & 0.7 & 0.6 \\
 1501-1510 & 0.4 & 0.6 \\
 1991-2000 & 0.5 & 0.7 \\
 
 \hline
 \end{tabular} 
 \end{center} 
\captionof{table}{Precision at 10 evaluated at different ranks of the unigram and $\alpha$-mix retrieval results  \label{ta:result} 
}
 
\end{minipage}

\end{table}

\textbf{Precision at k:} A second evaluation target is to test how dense the relevant documents on different ranges in the ranks are. The \textit{precision at k} measure can be utilized to determine the quality of the process by manually assessing the first 10 documents downwards from the ranks 1, 101, 501, 1001, 1501, 1991. Documents were marked as relevant in case a domain expert was able to annotate text snippets therein regarding arguments, topics or claims representing a discourse framed in neoliberal terminology.
The results in table \ref{ta:result} confirm the usefulness of our
 approach. Density of positively evaluated results in the upper ranks is very high and gets lower towards the bottom of the list. Precision in the best performing system remains high also in lower ranks while it drops off in comparison with the system which solely exploits unigram matchings.

\section{Conclusion}
We presented an extension of the VSM approach of IR by exploiting automatically extracted dictionaries and co-occurrence data from manually compiled reference collections. This method has proven to produce valuable results for Content Analysis studies to extract collections specific to a certain research interest from large unspecific corpora. The use of co-occurrence data improves upon merely taking into account raw frequencies. Results could be enhanced further by creating dictionaries with the help of Topic Models. The retrieval mechanism allows domain experts, such as social scientists, to build collections for further analyses based on paradigmatic example documents or theory texts representing abstract domain knowledge.



\bibliographystyle{splncs03}
\bibliography{eacl2014}








\end{document}